\documentclass[prd, 10pt,aps,nofootinbib, floatfix, notitlepage,twocolumn]{revtex4-1}
\usepackage{amsmath,amsfonts,amssymb}
\usepackage{mathrsfs}
\usepackage{multirow}
\usepackage{graphicx}
\usepackage[english]{babel} 
\usepackage{url} 
\usepackage{color}
\usepackage[utf8]{inputenc} 
\usepackage{float}
\usepackage[colorlinks=true,citecolor=blue,urlcolor=blue]{hyperref}
\usepackage{booktabs}
\usepackage{subfigure}

\usepackage[T1]{fontenc}
\begin{document}

%\preprint{}

%\title{Generalized dark matter with $\gamma$-DM interaction}
\title{What Prevents Resolving the Hubble Tension through Late-Time Expansion Modifications?}


\author{Zhihuan Zhou}
\email{zhizhuanzhou1@163.com}
\altaffiliation[Also at ]{}%Lines break automatically or can be forced with \\
\author{Zhuang Miao}%
\author{Sheng Bi}%
%\author{Xiaodun Deng}%
\author{Chaoqian Ai}%
%\email{acq0003@live.com}
\affiliation{%Institute of Theoretical Physics \\
	School of Engineering \\
	Xi'an International University \\
	Xi'an 710077, People's Republic of China}%
	
\author{Hongchao Zhang}
\email{zhanghongchao852@live.com}	
\affiliation{%Institute of Theoretical Physics \\
	Department of Physics \\
	Liaoning Normal University \\
	Dalian, 116029, People's Republic of China}%

\date{\today}

\begin{abstract}
We demonstrate that Type Ia supernovae (SNe Ia) observations impose the critical constraint for resolving the Hubble tension through late-time expansion modifications. Applying the Fisher-bias optimization framework to cosmic chronometers (CC), baryon acoustic oscillations (BAO) from DESI DR2, Planck CMB, and Pantheon+ data, we find that: (i) deformations in $H(z \lesssim 3)$ (via $w(z)$ reconstruction) can reconcile tensions between CC, Planck, DESI BAO, and SH0ES measurements while maintaining or improving fit quality ($\Delta\chi^2 < 0$ relative to $\Lambda$CDM); (ii) In the neighborhood of Planck best-fit $\Lambda$CDM model, no cosmologically viable solutions targeting $H_0 \gtrsim 69$ satisfy SNe Ia constraints. MCMC validation confirms the maximum achievable $H_0 = 69.09\pm0.30$  ($\chi^2_{\rm BF} \approx \chi^2_{\Lambda\rm CDM}$) across all data combinations, indicating that the conflict between late-time $w(z)$ modifications and SNe Ia observations prevents complete resolution of the Hubble tension.
\end{abstract}

%\keywords{}

\maketitle

\section{Introduction}

The $\Lambda$ Cold Dark Matter ($\Lambda$CDM) model has achieved remarkable success in describing the evolution of our Universe through precision fits to diverse cosmological observations \cite{SupernovaSearchTeam:1998fmf,Planck:2018vyg,Planck:2019nip,Abbott:2017wau,Alam:2016hwk}.  However, the nature of its dominant components—dark matter and dark energy (DE)—remains unknown, while growing evidence reveals internal tensions in its parameter space \cite{CosmoVerse:2025txj,Poulin:2023lkg,Verde:2023lmm, Verde:2019ivm,DiValentino:2021izs,Wong:2019kwg}. The most significant challenge emerges in measuring the Hubble constant $H_0$, which governs the Universe's current expansion rate.  A persistent $>5\sigma$ discrepancy persists between early-Universe inferences from the \textit{Planck} Cosmic Microwave Background (CMB) ($H_0 = 67.36 \pm 0.54$ under $\Lambda$CDM) \cite{Aghanim:2018eyx} and late-Universe measurements using Cepheid-calibrated Type Ia supernovae (SNe Ia) by SH0ES 2024 ($H_0 = 73.17 \pm 0.86$) \cite{Riess:2024vfa}. This ``Hubble tension'' represents a pivotal challenge for modern cosmology, potentially signaling physics beyond the standard model. The $\Lambda$CDM model also exhibits tension in the matter clustering amplitude $S_8 \equiv \sigma_8 \sqrt{\Omega_{m,0}/0.3}$, where late-universe probes report consistently lower values than CMB-derived predictions. Weak lensing surveys like DES-Y3\cite{DES:2021bvc} ($S_8 = 0.759^{+0.024}_{-0.021}$) and KiDS-1000\cite{KiDS:2020ghu} ($S_8 = 0.759^{+0.024}_{-0.021}$) disagree with the \textit{Planck}-2018 CMB value $S_8 = 0.834 \pm 0.016$ \cite{Planck:2018vyg} in  $\Lambda$CDM model at $2$--$3\sigma$ significance \cite{Kilo-DegreeSurvey:2023gfr,Dalal:2023olq, Chen:2024vvk,Abbott:2017wau,Hildebrandt:2016iqg}. While less severe than the Hubble tension, this $S_8$ discrepancy further motivates scrutiny of $\Lambda$CDM's assumptions about late-time structure growth \cite{Sabogal:2024yha}.

Early-Universe solutions modify either the pre-recombination expansion rate \cite{Poulin:2023lkg,Hill:2020osr,Velten:2021cqj,Vagnozzi:2021gjh} or ionization history \cite{Lee:2022gzh,Lee:2025yah,Mirpoorian:2025rfp,Jedamzik:2025cax,Mirpoorian:2024fka} to reduce the sound horizon scale ($r_d$). Although both CMB anisotropies and baryon acoustic oscillation (BAO) measurements precisely constrain the angular sound horizon scale -- each defining distinct degeneracy directions in the $r_d$-$H_0$ plane -- the slope discrepancy between these probes shows that simply decreasing $r_d$ cannot reconcile \textit{Planck} and SH0ES measurements without violating BAO or weak lensing constraints \cite{Hill:2020osr,Velten:2021cqj}.  %Recent work \cite{modified_recomb} demonstrates that modified recombination models can reduce the tension below $2\sigma$ while improving fits to CMB/BAO data and alleviating the $S_8$ tension, though complete resolution may require additional late-time effects.

Late-time solutions modifying the expansion history at $z\lesssim 3$ face their own challenges \cite{Benevento:2020fev,Efstathiou:2021ocp,Gomez-Valent:2023uof,Yang:2018qmz,Heisenberg:2022gqk,Zhou:2021xov,Escamilla:2023oce,Alestas:2021xes}. While CMB ($z\sim1100$) and BAO ($z\sim0.4$) both anchor the sound horizon scale, simultaneously increasing $H_0$ requires either a sharp phantom transition of DE equation of state (EoS) at $z\lesssim0.4$ or a late-time jump in the effective gravitational constant $G_{\rm eff}$ \cite{Alestas:2020zol,Marra:2021fvf} (at $z_t \lesssim 0.01$) to explain the $\Delta M_B \approx -0.2$ mag calibration offset.
While a drastic evolution of $w(z)$ in the late-universe is strongly disfavored by SNe Ia observations, this raises a critical question: Can any model, free from parametric assumptions, self-consistently reconcile (i) CMB, (ii) BAO, and (iii) calibrated SNe Ia? self-consistently reconcile (i) CMB, (ii) BAO, and (iii) Cepheid-calibrated SNe Ia?  Notably, the tension between BAO and calibrated SNe Ia persists even in scenarios with early-universe modifications, raising non-trivial questions about the existence of self-consistent solutions to the Hubble tension. It's argued that a promising way forward should ultimately involve a combination of early- and late-time new physics \cite{Vagnozzi:2023nrq}. 
Other viable solutions includes multi-interactions interactions \cite{Liu:2023kce,Becker:2020hzj,Yang:2019nhz,Vattis:2019efj,Li:2023fdk}, modified gravity \cite{Addazi:2021xuf, AlvesBatista:2023wqm},cosmic voids \cite{Jia:2025prq,Pisani:2019cvo} (see Ref. \cite{CosmoVerse:2025txj} for a recent review).

In this Letter, we transcend the conventional model-specific approaches to resolving the Hubble tension by employing the Fisher-bias formalism \cite{Lee:2022gzh,Lee:2025yah} to late expansion rate modifications. This framework systematically identifies minimal, observationally grounded extensions to $\Lambda$CDM that generate precise cosmological parameter shifts without compromising fit quality. While established sampling methods like MCMC offer rigorous parameter estimation, their computational demands grow prohibitive in high-dimensional function spaces. Our approach circumvents this limitation by combining analytic gradient calculations with efficient exploration of continuous dark energy equation-of-state $w(z)$ modifications. This approach bridges the gap between purely data-driven methods \cite{Gerardi:2019obr,Hart:2019gvj} and traditional techniques, preserving physical interpretability while maintaining the statistical reliability.

Building upon the latest baryon acoustic oscillation (BAO) measurements from DESI Data Release 2 \cite{DESI:2025zgx}, we re-examine the persistent tension between CMB, late-time BAO, and Type Ia supernova observations. While previous studies have established the challenges of reconciling these datasets through late-time modifications \cite{Benevento:2020fev,Gomez-Valent:2023uof,Heisenberg:2022gqk}, our Fisher-bias approach provides new insights by systematically quantifying both the necessary departures from $\Lambda$CDM and the specific redshift ranges where expansion history modifications yield optimal consistency improvements. This framework enables us to distinguishing whether tensions arise from collective inconsistencies across all datasets or specific conflicts between particular probes, and assess whether current tensions reflect fundamental limitations of late-time solutions or can be resolved through more flexible cosmological parameterizations.

This work is organized as follows. Section~\ref{sec:method} develops the Fisher-bias formalism for late-time solutions to the Hubble tension. Section~\ref{sec:data} details the observational datasets and covariance modeling. Section~\ref{sec:discussion} presents our key results on the required expansion history modifications. We conclude in Section~\ref{sec:conclusions} by outlining future directions.
\\

\section{Methodology}\label{sec:method}
\subsection{Fisher-Bias Formalism and Hubble Optimization}
We employ the Fisher-bias formalism \cite{Lee:2022gzh} to analyze late-time modifications resolving the Hubble tension within the cosmological parameter space $\vec{\Omega} = \{\omega_c, \omega_b, H_0, \tau, \ln(10^{10}A_s), n_s\}$. Theoretical predictions $\mathbf{X}(\vec{\Omega})$ are compared to observations $\mathbf{X}^{\mathrm{obs}}$ through the $\chi^2$ statistic:
\begin{equation}
	\chi^2(\vec{\Omega}) \equiv [\bm{X}(\vec{\Omega}) - \bm{X}^{\rm obs}] \cdot \bm{M} \cdot [\bm{X}(\vec{\Omega}) - \bm{X}^{\rm obs}],
\end{equation}
with $\bm{M} = \bm{\Sigma}^{-1}$ the inverse covariance matrix. 

We model DE as an effective fluid with perturbations evolving according to \cite{Ballesteros:2010ks,Ma:1995ey}. The effective sound speed are fixed at $\hat{c}^2_s\equiv \delta p/\delta\rho=1$ so that the fluid can not be clustered.
Our analysis incorporates perturbations to the DE EoS through the transformation $w(z) \to w(z) + \Delta w(z)$. Rather than directly minimizing $\Delta w(z)$, we optimize these perturbations through their impact on the Hubble parameter $\Delta H(z)$. This approach is motivated by three key considerations:

\begin{enumerate}
    \item \textbf{Observational robustness}: $H(z)$ is directly constrained by cosmic chronometers and BAO measurements, whereas $w(z)$ requires model-dependent integration of the Friedmann equations.

	\item \textbf{Theoretical stability}: The linear response of the Hubble parameter naturally suppresses high-frequency oscillations in $\Delta w(z)$, ensuring physically plausible solutions.
	
	\item \textbf{Energy conservation}: $H(z)$ perturbations automatically satisfy  the continuity equation: $\dot{\rho} + 3H(\rho + P) = 0$, and naturally bounds sound speed $c_s^2 = w - \dot{w}/[3H(1+w)]$ avoiding divergences in direct $w(z)$ optimization..
\end{enumerate}

Our goal is to find the smallest perturbation $\Delta H(z)$ that shifts the best-fit Hubble constant to the SH0ES target $H_0^{\rm target} = 73.0$ while preserving fit quality relative to $\Lambda$CDM. Formally, we solve the constrained optimization:
\begin{equation}
	\textrm{minimize}(|| \Delta H||^2) \quad \textrm{subject to} \quad 
	\begin{cases} 
		\vec{\Omega}_{\rm BF}[\Delta w] = \vec{\Omega}_{\rm target}, \\
		\Delta \chi^2_{\rm BF}[\Delta w] \leq 0,
	\end{cases}
	\label{eq:optimization}
\end{equation}
with $|| \Delta H||^2 \equiv \int dz \left[\Delta H(z)\right]^2$ quantifies deviations from the $\Lambda$CDM expansion history, and $\Delta \chi^2 \equiv \chi^2(\Delta w) - \chi^2_{\Lambda\rm CDM}$ ensures no degradation in fit quality.

\subsection{Functional Expansion and Response Theory}

To parametrize deviations from $\Lambda$CDM ($w=-1$), we model the DE EoS perturbations using Gaussian basis functions:
\begin{equation}
	\Delta w(z) = \sum_{j=1}^{N} c_j \exp\left(-\frac{(z - z_j)^2}{2\sigma_j^2}\right),\label{eq:delta_w}
\end{equation}
with $N$ equally spaced nodes $z_j \in [0, 2]$ and widths $\sigma_j = \Delta z/(2\sqrt{2\ln 2})$ where $\Delta z = 2/(N-1)$ is the node spacing. 
The response of cosmological parameters to perturbations is derived through functional derivatives:
\begin{align}
	\Delta \Omega_{\rm BF}^i &= \int dz\; \mathcal{R}^i_{\Omega}(z) \Delta w(z),
	\label{eq:DObf_X} \\
	\Delta \chi_{\rm BF}^2 &= \int dz\; \mathcal{R}_{\chi}(z) \Delta w(z) \notag \\
	&+ \frac{1}{2} \iint dz dz'\; \mathcal{R}_{\chi}^{(2)}(z,z') \Delta w(z) \Delta w(z'),
\end{align}
where response kernels are:
\begin{align}
	\mathcal{R}^i_{\Omega}(z) &\equiv -(F^{-1})_{ij} \frac{\partial \bm{X}}{\partial \Omega^j} \cdot \bm{M}\cdot\frac{\delta \bm{X}}{\delta w(z)},
	\\
	\mathcal{R}_{\chi}(z) &\equiv 2[\bm{X}_{\rm fid} - \bm{X}^{\rm obs}] \cdot\widetilde{\bm{M}} \cdot \frac{\delta \bm{X}}{\delta w(z)},
	\\
	\mathcal{R}_{\chi}^{(2)}(z,z') &\equiv 2 \frac{\delta \bm{X}}{\delta w(z)}\cdot\widetilde{\bm{M}}\cdot\frac{\delta \bm{X}}{\delta w(z')},
\end{align}
with $\widetilde{\bm{M}}$ the marginalized inverse covariance:
\begin{equation}
	\widetilde{M}_{\alpha\beta} = M_{\alpha\beta} - M_{\alpha \gamma} \frac{\partial X^\gamma}{\partial \Omega^i}(F^{-1})_{ij}\frac{\partial X^\sigma}{\partial \Omega^j}M_{\sigma\beta}.
\end{equation}

\subsection{Hubble Perturbation Minimization}

The Hubble perturbation functional is discretized as:
\begin{equation}
	\Delta H(z) = \int dz' \frac{\delta H(z)}{\delta w(z')} \Delta w(z') \approx \sum_j c_j \mathcal{K}_j(z),
\end{equation}
where the kernel components are:
\begin{equation}
	\mathcal{K}_j(z) \equiv \int dz' \frac{\delta H(z)}{\delta w(z')} \phi_j(z').
	\label{eq:kernel}
\end{equation}
The minimization target becomes:
\begin{equation}
	||\Delta H||^2 = \bm{c}^T \bm{\mathcal{Q}} \bm{c}, \quad \mathcal{Q}_{mn} \equiv \int dz  \mathcal{K}_m(z) \mathcal{K}_n(z),
\end{equation}
%which automatically suppresses unphysical oscillations through the causal structure of the functional derivative:
%\begin{equation}
%	\frac{\delta H(z)}{\delta w(z')} = \begin{cases} 
%		\dfrac{3H_0^2 \Omega_\Lambda}{2H(z)} \dfrac{e^{3\int_0^{z'} \frac{1+w}{1+\zeta}d\zeta}}{1+z'} & z' < z \\
%		0 & z' > z 
%	\end{cases}.
%\end{equation}
The complete optimization problem is thus reduced to quadratic programming:
\begin{equation}
	\textrm{minimize}_{\bm{c}} \left( \bm{c}^T \bm{\mathcal{Q}} \bm{c} \right) \quad \textrm{subject to} \quad 
	\begin{cases}
		\bm{\mathcal{R}}_{\Omega} \bm{c} = \Delta\vec{\Omega}_{\rm target} \\
		\bm{c}^T \bm{\mathcal{R}}_{\chi}^{(2)} \bm{c} + \bm{\mathcal{R}}_{\chi}^T \bm{c} \leq 0
	\end{cases}
\end{equation}
with response matrices constructed from Eqs.~\eqref{eq:DObf_X}-\eqref{eq:kernel}. %This formulation ensures solutions are theoretically consistent and computationally tractable.

\begin{figure}
	\begin{center}
		\includegraphics[scale = 0.34]{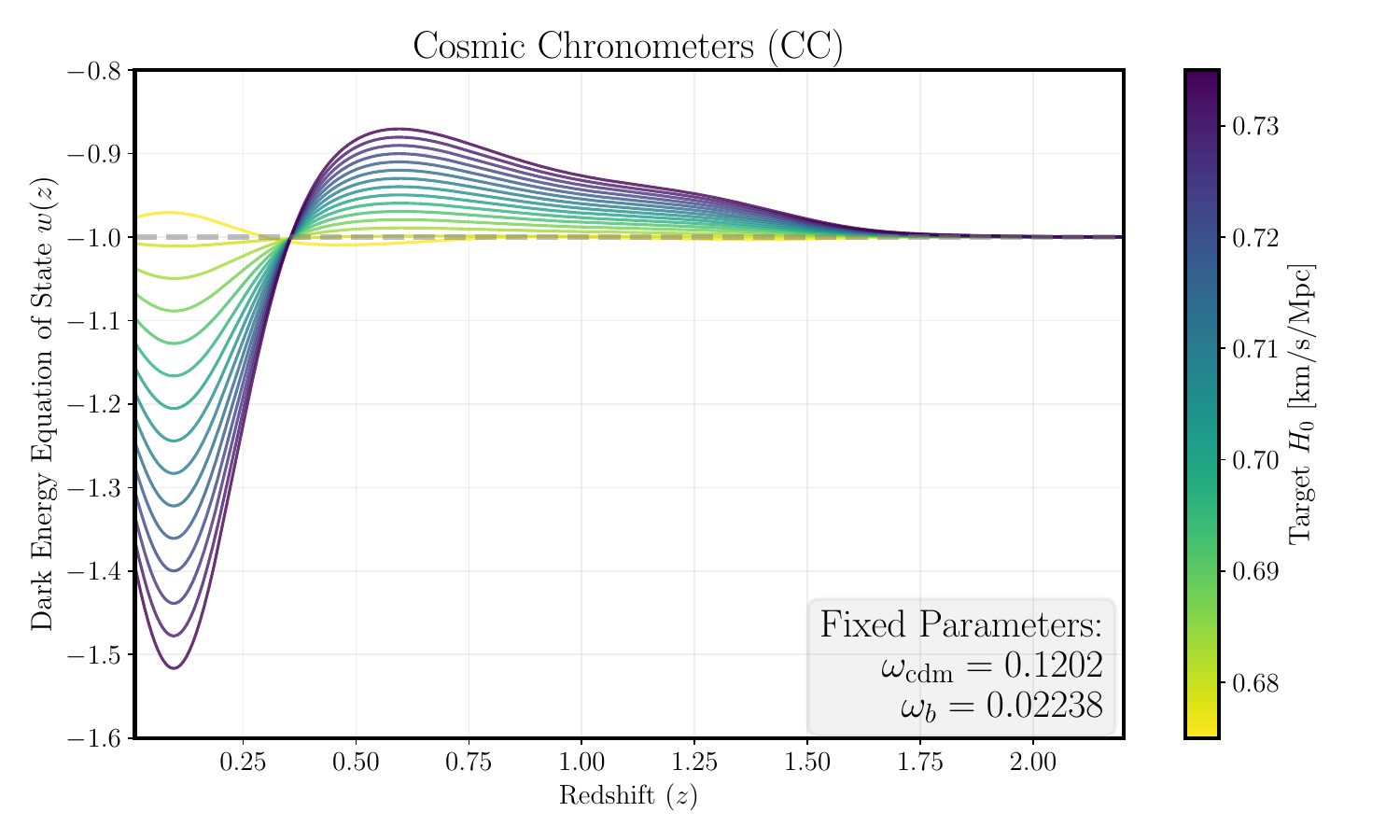}
		\includegraphics[scale = 0.34]{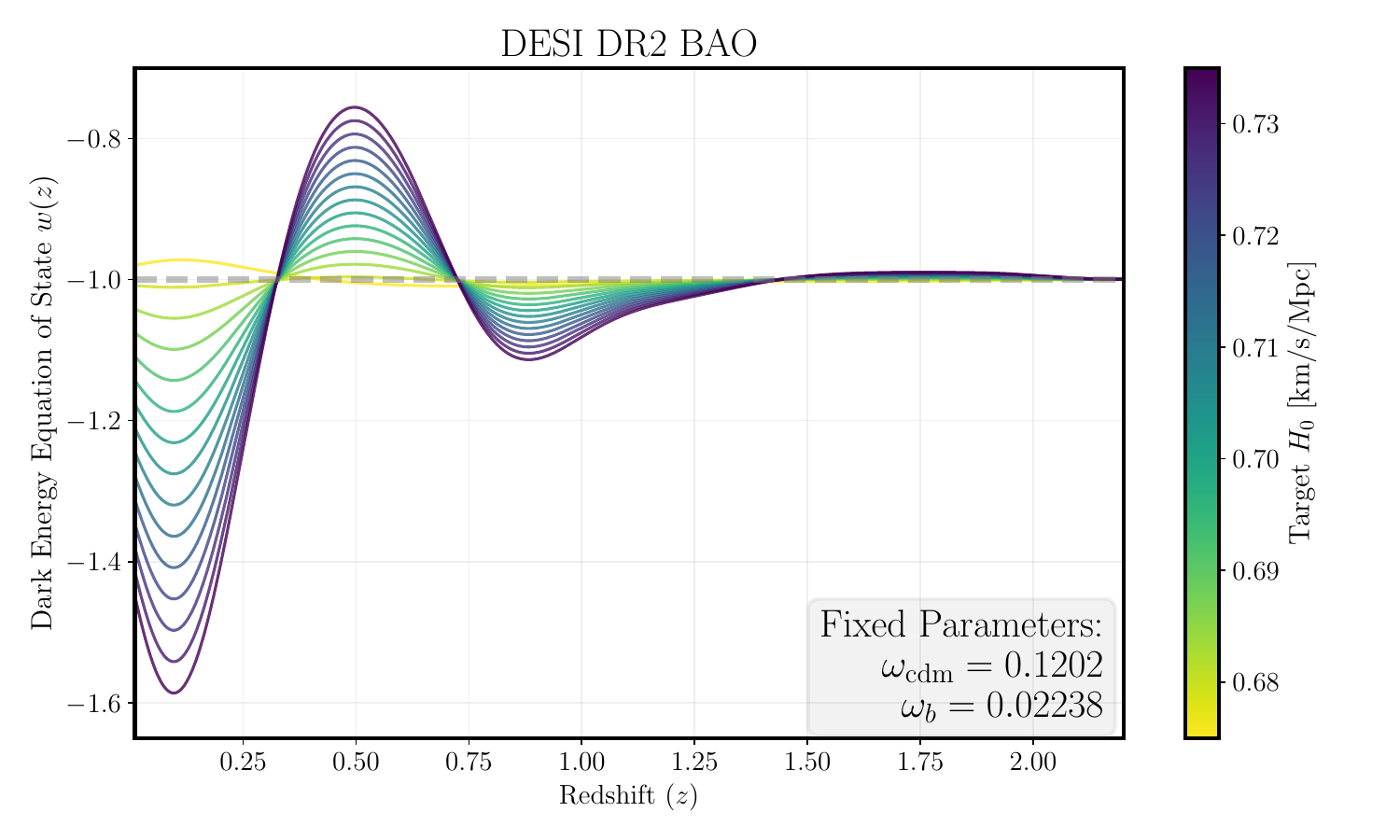}
		\includegraphics[scale = 0.34]{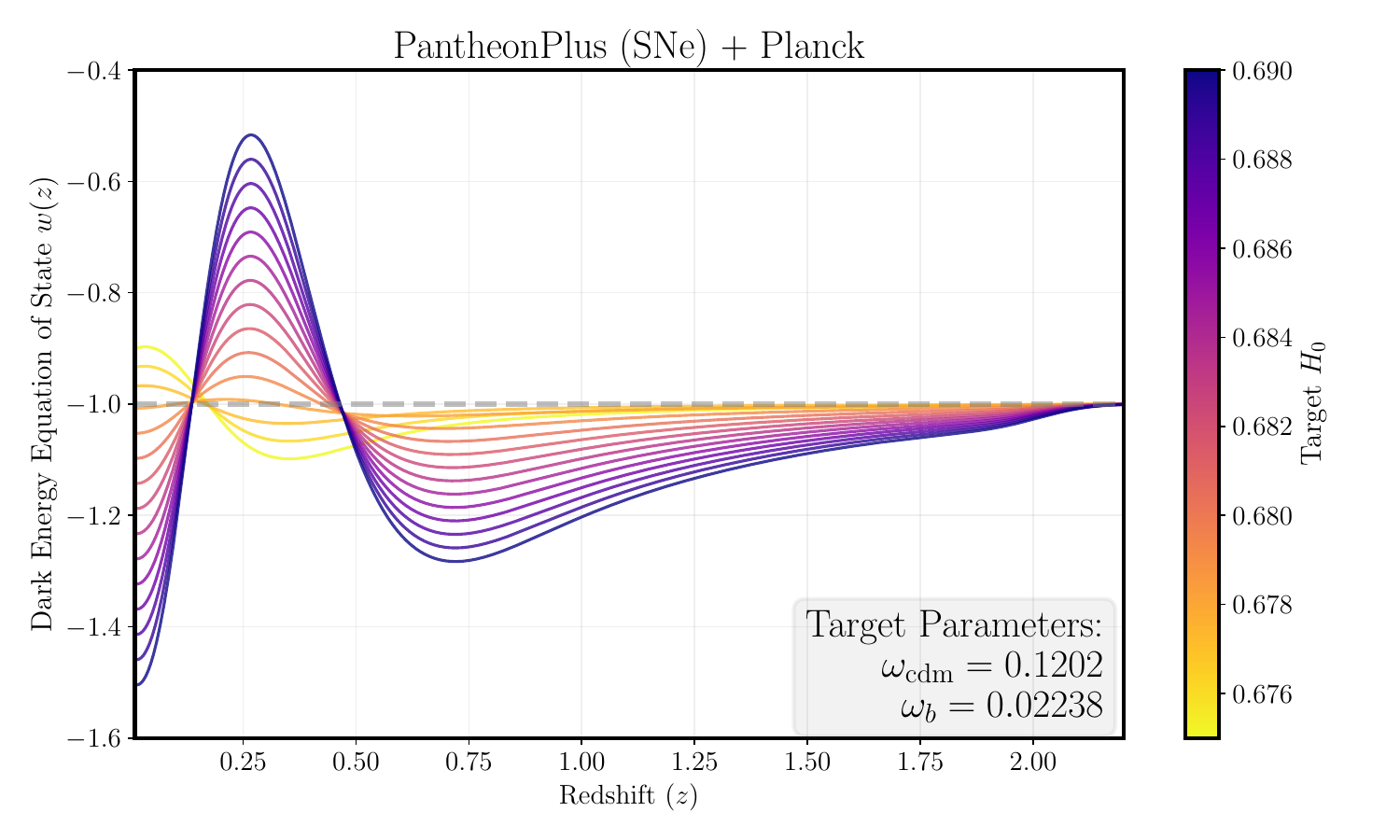}
		\includegraphics[scale = 0.34]{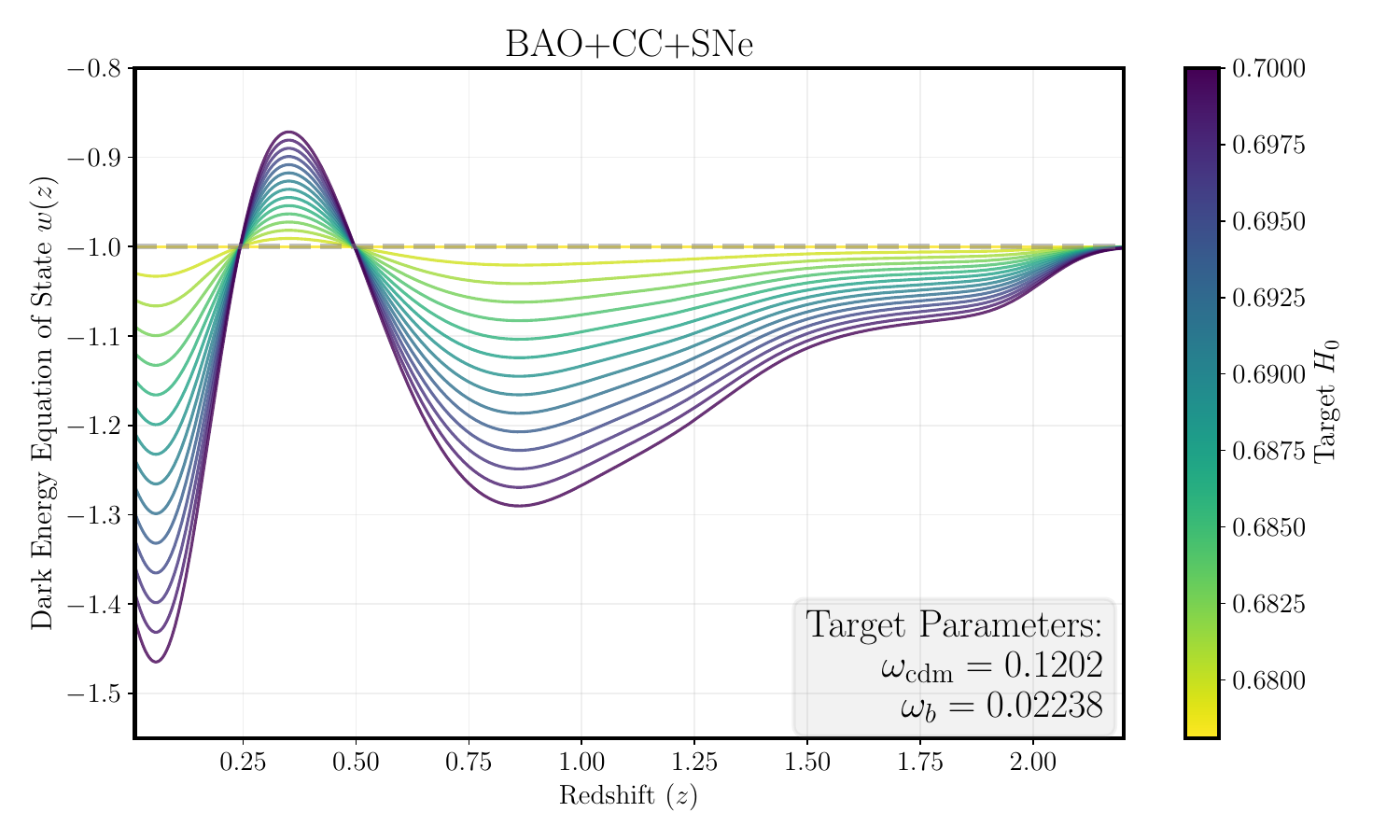}
	\end{center}
	\caption{Solutions for $w(z\lesssim 2)$ given target value of $H_0$. The vertical panels show solutions for: (a) cosmic chronometers (CC) data with maximum target $H_0 = 73.17$; (b) DESI BAO measurements ($H_0 = 73.17$); (c) Pantheon+ SNe combined with Planck distance prior ($H_0 = 69.0$); and (d) the full combination of BAO+CC+SNe data with target $H_0 = 69.0$). All solutions preserve the Planck $\Lambda$CDM best-fit parameters.
	}\label{fig:w_fld_compare}
\end{figure}

\begin{figure}
	\begin{center}
		\includegraphics[scale = 0.40]{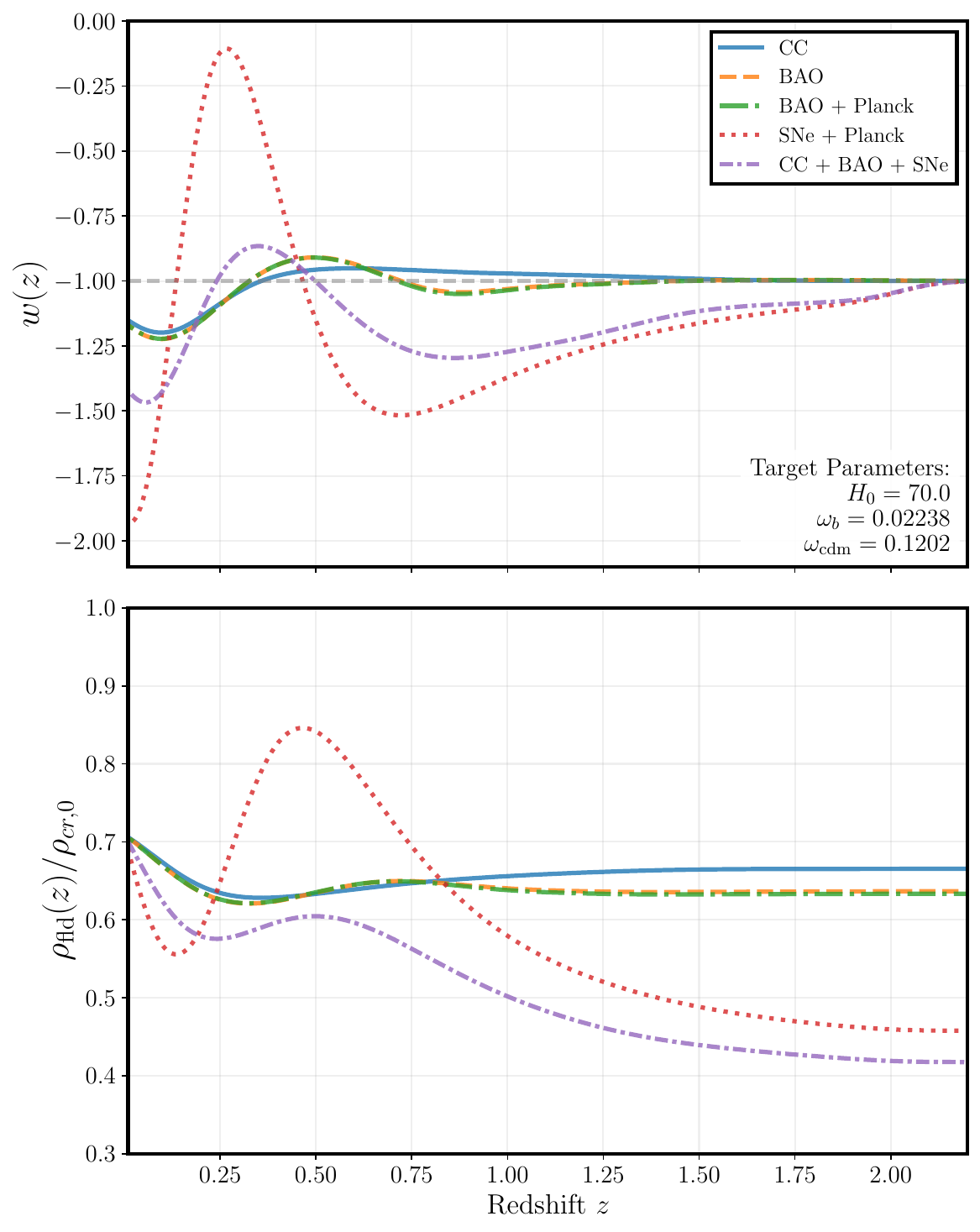}
	\end{center}
	\caption{\textit{Upper panel}: The DE EoS $w(z)$ showing solutions from cosmic chronometers (CC) alone (blue), DESI BAO (orange), BAO+Planck distance prior (PLC)+SNe (green), and the full combination CC+BAO+SNe (red). \textit{Lower panel}: Corresponding evolution of the relative DE density $\rho_{\rm fld}(z)/\rho_{\rm crit,0}$ for each case. All solutions with targeting best-fit parameters ($H_0 = 70.0$, $\omega_b = 0.02237$, $\omega_{\rm cdm} = 0.1200$) while optimizing the late-time expansion history through Fisher-bias analysis.
	}\label{fig:DE_evolution_comparison}
\end{figure}

\section{STATISTICAL METHODOLOGY AND DATASETS}\label{sec:data}
We implement the GDE scenario as modifications to
the publicly available Einstein-Boltzmann code \href{http://class-code.net/}{CLASS} \cite{Lesgourgues:2011re,Blas:2011rf} package.
The non-linear matter power spectrum required by redshift-space distortion (RSD) likelihoods are computed using the ``HMcode'' \cite{Mead:2015yca,Mead:2016ybv,Mead:2020vgs} implemented in CLASS.
The MCMC analyses are performed 
using the publicly available code \href{https://cobaya.readthedocs.io}{COBAYA} \cite{Torrado:2020dgo} package
with a Gelman-Rubin \cite{10.1214/ss/1177011136} convergence
criterion $R-1 < 0.05$. The plots have been obtained
using the \href{https://getdist.readthedocs.io}{\texttt{GetDist}} \cite{Lewis:2019xzd} package
We utilize multiple cosmological probes spanning different redshifts and physical scales to constrain late-time expansion history modifications, combining DESI BAO, Pantheon+ SNe Ia, cosmic chronometers, and Planck distance priors with their full covariance matrices.

\subsection{DESI BAO (DR2)}
\label{subsec:desi_dr2}

The Dark Energy Spectroscopic Instrument (DESI) Data Release 2 \cite{DESI:2025zgx} provides BAO measurements using 14 million extragalactic objects across four distinct tracer classes. The BAO measurements are reported in nine redshift bins spanning $0.295 \leq z \leq 2.330$ as:
\begin{equation}
	\left( \frac{D_M(z)}{r_d}, \frac{D_H(z)}{r_d} \right),
\end{equation}
where $D_M(z)$ is the transverse comoving distance, $D_H(z) = c/H(z)$ is the Hubble distance, and $r_d$ is the sound horizon at the drag epoch. We incorporate the full non-diagonal covariance matrix accounting for cross-correlations between redshift bins and tracer types.

\subsection{Pantheon+ Supernovae}
\label{subsec:panthon_plus}
The Pantheon+ sample \cite{Scolnic:2021amr, Brout:2022vxf} comprises 1701 light curves from 1550 distinct Type Ia supernovae. To avoid calibration systematics associated with the Hubble tension. The Dark Energy Survey (DES) Year 5 data release includes a new homogeneous sample of 1,635 photometrically classified Type Ia supernovae spanning the redshift range $0.1 < z < 1.3$ \cite{DES:2024jxu}. We have excluded the SH0ES prior In the Fisher-Bias framework, and the absolute magnitude $M_B$ is analytically marginalized through the modified covariance matrix:

\begin{equation}
	\widetilde{\mathbf{C}}^{-1} = \mathbf{C}^{-1} - \frac{\mathbf{C}^{-1}\mathbf{1}\mathbf{1}^T\mathbf{C}^{-1}}{\mathbf{1}^T\mathbf{C}^{-1}\mathbf{1}}
\end{equation}
where $\mathbf{1}$ denotes the unit vector. This preserves the full statistical and systematic covariance $\mathbf{C}$ while removing the $M_B$ dependence. We apply a conservative $z<0.1$ cut to minimize peculiar velocity systematics.

\subsection{Cosmic Chronometers}
\label{subsec:cc}
Cosmic chronometer measurements directly constrain the Hubble parameter through differential aging of passively evolving galaxies. We utilize 32 measurements summarized in Ref. \cite{Wu:2025wyk} spanning $0.07 < z < 1.97$ (See Table \ref{tab:cc}). The covariance matrix accounts for systematic uncertainties in stellar population synthesis models.

\begin{table}[t]
	\renewcommand{\arraystretch}{0.5}
	\setlength{\tabcolsep}{5pt}
	\centering
	\caption{The 32 $H(z)$ measurements obtained with the CC method.}
	\label{CCHz}
	\begin{tabular}{|l|l|l|l|}
		\hline
		Redshift $z$ & $H(z)\ [\,\rm km/s/Mpc\,]$ & Reference \\
		\hline
		\hline
		0.07 & $69.0 \pm 19.6$ & \cite{Zhang:2012mp} \\
		0.09 & $69 \pm 12$ & \cite{Simon:2004tf} \\
		0.12 & $68.6 \pm 26.2$ & \cite{Zhang:2012mp} \\
		0.17 & $83 \pm 8$ & \cite{Simon:2004tf} \\
		0.179 & $75 \pm 4$ & \cite{Moresco:2012jh} \\
		0.199 & $75 \pm 5$ & \cite{Moresco:2012jh} \\
		0.2 & $72.9 \pm 29.6$ & \cite{Zhang:2012mp} \\
		0.27 & $77 \pm 14$ & \cite{Simon:2004tf} \\
		0.28 & $88.8 \pm 36.6$ & \cite{Zhang:2012mp} \\
		0.352 & $83 \pm 14$ & \cite{Moresco:2012jh} \\
		0.38 & $83 \pm 13.5$ & \cite{Moresco:2016mzx} \\
		0.4 & $95 \pm 17$ & \cite{Simon:2004tf} \\
		0.4004 & $77 \pm 10.2$ & \cite{Moresco:2016mzx} \\
		0.425 & $87.1 \pm 11.2$ & \cite{Moresco:2016mzx} \\
		0.445 & $92.8 \pm 12.9$ & \cite{Moresco:2016mzx} \\
		0.47 & $89 \pm 49.6$ & \cite{Ratsimbazafy:2017vga} \\
		0.4783 & $80.9 \pm 9$ & \cite{Moresco:2016mzx} \\
		0.48 & $97 \pm 62$ & \cite{Stern:2009ep} \\
		0.593 & $104 \pm 13$ & \cite{Moresco:2012jh} \\
		0.68 & $92 \pm 8$ & \cite{Moresco:2012jh} \\
		0.75 & $98.8 \pm 33.6$ & \cite{Borghi:2021rft} \\
		0.781 & $105 \pm 12$ & \cite{Moresco:2012jh} \\
		0.875 & $125 \pm 17$ & \cite{Moresco:2012jh} \\
		0.88 & $90 \pm 40$ & \cite{Stern:2009ep} \\
		0.9 & $117 \pm 23$ & \cite{Simon:2004tf} \\
		1.037 & $154 \pm 20$ & \cite{Moresco:2012jh} \\
		1.3 & $168 \pm 17$ & \cite{Simon:2004tf} \\
		1.363 & $160 \pm 33.6$ & \cite{Moresco:2015cya} \\
		1.43 & $177 \pm 18$ & \cite{Simon:2004tf} \\
		1.53 & $140 \pm 14$ & \cite{Simon:2004tf} \\
		1.75 & $202 \pm 40$ & \cite{Simon:2004tf} \\
		1.965 & $186.5 \pm 50.4$ & \cite{Moresco:2015cya} \\
		\hline
	\end{tabular}\label{tab:cc}
\end{table}

\subsection{Planck datasets}
\label{subsec:planck_full}
We consider the CMB distance prior derived from final Planck 2018 release\cite{Chen:2018dbv} in Fisher-Bias analysis. These priors include the shift parameter $\mathcal{R}$, the acoustic scale $\ell_A$, and the baryon density $\Omega_b h^2$.
For MCMC validation of solutions,  we employ the \emph{Planck} 2018 low-$\ell$ TT+EE and \emph{Planck} 2018 high-$\ell$ TT+TE+EE temperature and polarization power spectrum \cite{Planck:2019nip, Planck:2018lbu}. %To marginalize over nuisance parameters, we use the ``lite'' likelihoods. 

\begin{figure*}
	\begin{center}
		\includegraphics[scale = 0.37]{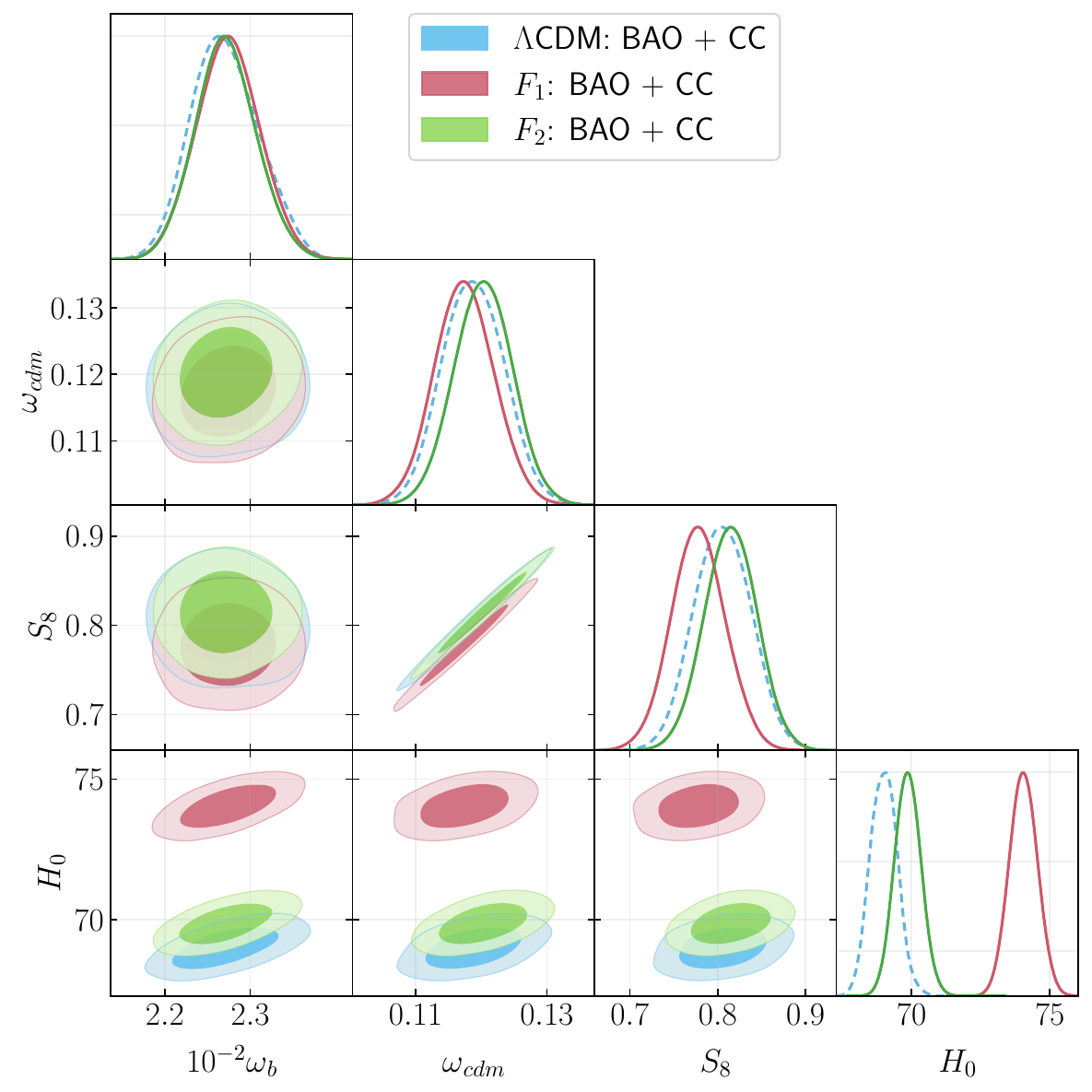}
		\includegraphics[scale = 0.37]{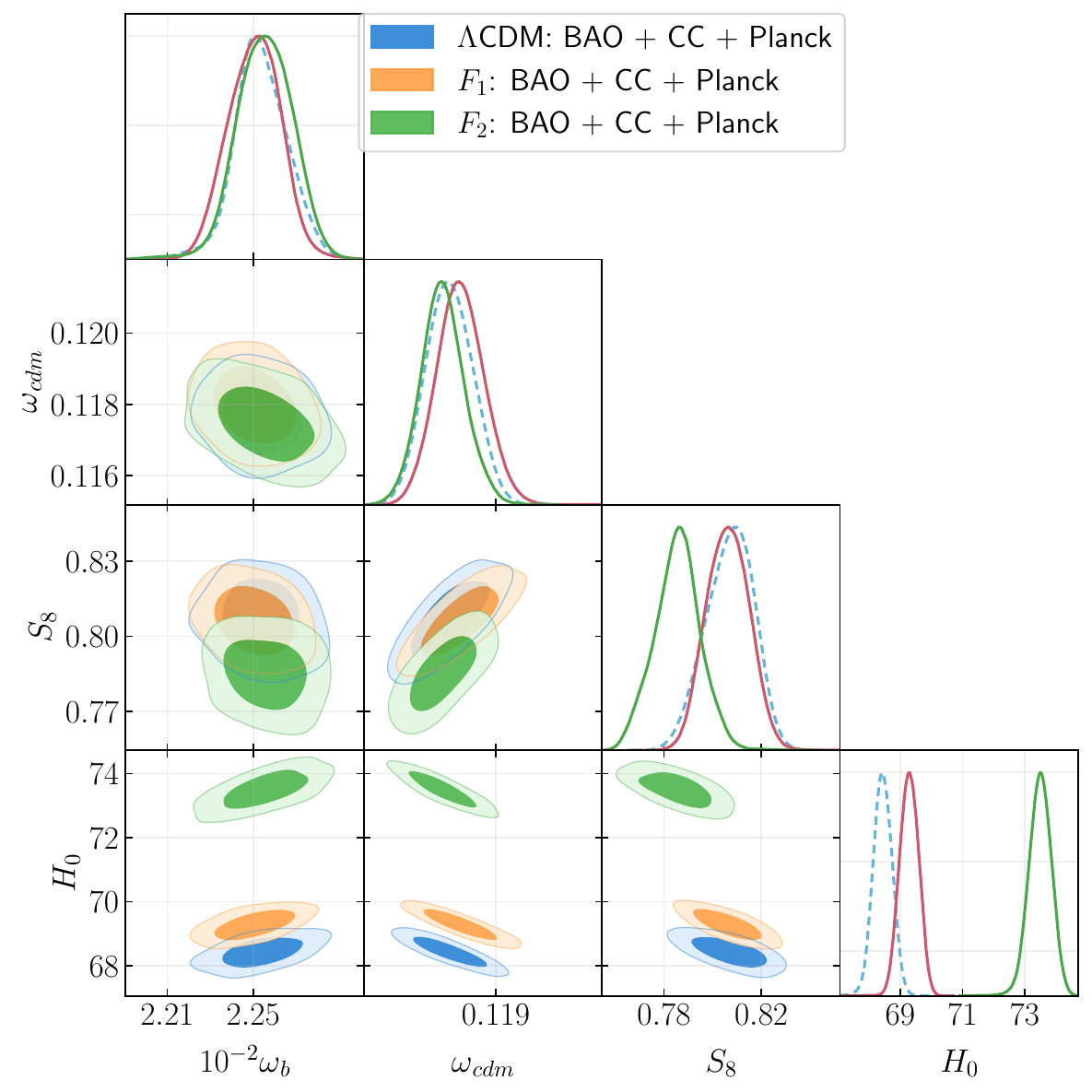}
	\end{center}
	\caption{Constraints on cosmological parameters for the three compared models (F1, F2, $\Lambda$CDM) from different data combinations. The upper panel displays constraints from DESI BAO and cosmic chronometers (CC) data, while the lower panel incorporates additional Planck distance prior (PLC) measurements. All contours show $1\sigma$ and $2\sigma$ confidence regions. The Fisher-bias optimized models target different Hubble constant values: F1-BAO for $H_0 = 73.0$ and F2-SN for $H_0 = 69.0$.
	}\label{fig:compare_contour}
\end{figure*}

\section{RESULTS AND DISCUSSIONS}\label{sec:discussion}

\label{subsec:universal_phantom}
The observed $w(z)$ behavior exhibits distinct physical patterns across different scenarios, as shown in Fig.~\ref{fig:w_fld_compare}. Both cosmic chronometers (CC) and BAO measurements independently reveal a consistent phantom crossing at $z\approx 0.3$ when targeting $H_0=73.0$. This transition displays three characteristic phases:
\begin{itemize}
	\item Quintessence-like behavior at $z \sim 0.5$
	\item Phantom crossing ($w = -1$) near $z \approx 0.3$ 
	\item Strong phantom phase ($w \approx -1.5$) at $z \sim 0.1$
\end{itemize}
The transition redshift $z_t \approx 0.3$ coincides with DE domination ($\Omega_{\Lambda}(z_t) \sim 0.5$), with this universal behavior across geometrically distinct probes suggesting new physics beyond $\Lambda$CDM.

\begin{table*}[t]
	\renewcommand{\arraystretch}{1.4}
	\setlength{\tabcolsep}{5pt}
	\centering
	\caption{Cosmological parameter constraints for F1 (BAO-derived), F2 (SNe-derived), and $\Lambda$CDM models across different data combinations. All values show the mean $\pm1\sigma$ confidence intervals.}
	\label{tab:combined_results}
	\begin{tabular}{|l|l|cccccc|c|}
		\hline
		\hline
		Dataset & Model & $10^{-2}\omega_b$ & $\omega_{\mathrm{cdm}}$ & $H_0$ & $S_8$ & $\Omega_{m,0}$ & $\chi^2$ \\
		\hline
		\hline
		\multirow{3}{*}{BAO+CC} 
		& F1 & $2.274\pm0.037$ & $0.1174\pm0.0045$ & $74.05\pm0.50$ & $0.778\pm0.031$ & $0.2568\pm0.0076$ & 11.64 \\
		& F2 & $2.271^{+0.036}_{-0.043}$ & $0.1188\pm0.0047$ & $69.00\pm0.48$ & $0.806\pm0.032$ & $0.2985\pm0.0087$ & 12.65 \\
		& $\Lambda$CDM & $2.271\pm0.036$ & $0.1203\pm0.0044$ & $69.87\pm0.48$ & $0.815\pm0.030$ & $0.2943\pm0.0079$ & 12.71 \\
		\hline
		\multirow{3}{*}{\emph{Planck}} 
		& F1 & $2.233\pm0.017$ & $0.1201^{+0.0017}_{-0.0015}$ & $72.76^{+0.81}_{-0.90}$ & $0.814^{+0.020}_{-0.018}$ & $0.2705\pm0.0086$ & 502.37 \\
		& F2 & $2.236^{+0.017}_{-0.023}$ & $0.1200^{+0.0019}_{-0.0017}$ & $68.33\pm0.83$ & $0.827^{+0.023}_{-0.021}$ & $0.306\pm0.011$ & 502.67 \\
		& $\Lambda$CDM & $2.236\pm0.018$ & $0.1197\pm0.0015$ & $67.47\pm0.70$ & $0.829^{+0.017}_{-0.019}$ & $0.3137\pm 0.0096$ & 502.58 \\
		\hline
		\multirow{3}{*}{BAO+CC+\emph{Plc}} 
		& F1 & $2.240^{+0.014}_{-0.017}$ & $0.11938^{+0.00079}_{-0.00064}$ & $72.51^{+0.35}_{-0.44}$ & $0.8091\pm0.0081$ & $0.2710^{+0.0045}_{-0.0036}$ & 516.48 \\
		& F2 & $2.248\pm0.013$ & $0.1183\pm0.0008$ & $69.15\pm0.32$ & $0.811\pm0.010$ & $0.2958\pm0.0041$ & 516.56 \\
		& $\Lambda$CDM & $2.2523^{+0.0087}_{-0.010}$ & $0.1178^{+0.0006}_{-0.0005}$ & $68.34^{+0.14}_{-0.22}$ & $0.8105\pm0.0087$ & $0.3019^{+0.0029}_{-0.0021}$ & 517.52 \\
		\hline
		\multirow{3}{*}{BAO+CC+\emph{Plc}+SN} 
		& F1 & $2.255\pm0.013$ & $0.1175^{+0.0005}_{-0.0006}$ & $73.50\pm0.36$ & $0.7856\pm0.0096$ & $0.2604\pm0.0035$ & 1287.05 \\
		& F2 & $2.251\pm0.012$ & $0.1180\pm0.0007$ & $69.09\pm0.30$ & $0.8063\pm0.0087$ & $0.2940\pm0.0038$ & 1224.73 \\
		& $\Lambda$CDM & $2.254\pm0.012$ & $0.1177\pm0.0007$ & $68.44\pm0.29$ & $0.807^{+0.011}_{-0.0089}$ & $0.3007\pm0.0038$ & 1224.65 \\
		\hline
	\end{tabular}
\end{table*}

\subsection{Resolution of CMB-BAO-SH0ES Tension}
\label{subsec:tension_resolution}

The optimized $w(z)$ perturbations derived from BAO datasets targeting $H_0 = 73.0$  effectively reconcile the tension between DESI DR2 BAO observations and SH0ES measurements. Crucially, this solution simultaneously elevates the \emph{Planck}-only derived Hubble constant to $H_0 = 72.76^{+0.81}_{-0.90}$ (see Table \ref{tab:combined_results}) through a phantom transition that proportionally modifies angular diameter distances across both BAO ($z\sim0$--$0.4$) and last-scattering ($z\sim1100$) epochs\footnote{The sound horizon scale $r_s$ remains invariant under late-time modifications, preserving the characteristic CMB angular power spectrum. Peak location shifts are precisely compensated by $H_0$ adjustments, as demonstrated in Ref.~\cite{Zhou:2021xov}.}.  %Such modification preserves the critical ratio $D_A(z_{\rm LSS})/D_A(z_*)$ while accommodating higher $H_0$ values. 
Notably, the $w(z)$ evolution obtained from BAO+PLC Fisher-bias analysis is consistent with the BAO-only reconstruction (see upper panel of Fig.~\ref{fig:DE_evolution_comparison}). Indicating that, the apparent tensions between cosmic chronometers, \emph{Planck} data, DESI BAO DR2, and SH0ES measurements can be fully resolved through late-time expansion rate modifications. 

A rapid consistency check can be performed in \href{http://class-code.net/}{CLASS} by fixing the angular acoustic scale $\theta_* = 1.0411 \times 10^{-2}$ (the Planck best-fit value) and employing the shooting method to determine the corresponding $H_0$. This yields $H_0 = 73.34$.  For rigorous validation, we perform MCMC analysis with a full  Planck TTTEEE+lensing likelihood (See Section \ref{sec:mcmc} for detailed discussion).

\subsection{Conflicts with Pantheon+ SNe Ia}
The joint analysis of Pantheon+ SNe Ia with either \emph{Planck} distance priors or DESI BAO measurements reveals fundamental limitations in reconstructing viable cosmological solutions when targeting $H_0 \gtrsim 69$. The analysis employs the marginalized Pantheon+ likelihood, which requires specification of a distance anchor - either the \emph{Planck}-determined sound horizon angle $\theta_*$ at $z\sim1100$ or DESI BAO measurements at $0.3 \leq z \leq 2.0$. Crucially, neither anchoring choice permits solutions simultaneously satisfying $H_0 > 69$ and the optimization constraint $\Delta \chi^2 \leq 0$ relative to $\Lambda$CDM.

As the target parameters approach this boundary ($H_0 \sim 69.0$), the reconstructed $w(z)$ develops pronounced low-redshift features: (1) an extreme phantom phase ($w(z) \ll -1$) at $z \sim 0.1$ and (2) significant DE density perturbations in the local universe (Fig.~\ref{fig:DE_evolution_comparison}). This pathological behavior suggests either a transition in the effective gravitational constant $G_{\rm eff}$ at $z<0.1$ \cite{Alestas:2020zol,Marra:2021fvf}, or Fundamental limitations of the reconstruction framework.

The tension structure reveals a critical pattern: while BAO and CMB measurements can be mathematically reconciled with local $H_0$ determinations through carefully tuned $w(z)$ oscillations, these solutions invariably violate Pantheon+ constraints. This demonstrates that the primary inconsistency lies not between SNe, BAO and CMB collectively, but specifically between supernova distances and other cosmological probes. The Pantheon+ dataset thus imposes unique restrictions on late-time expansion history modifications that cannot be circumvented through DE equation-of-state perturbations alone.

\subsection{MCMC Validation of Fisher-Bias Solutions}\label{sec:mcmc}
The Fisher-bias derived $w(z)$ profiles are validated through full MCMC analyses of two representative models
\begin{itemize}
	\item \textbf{F1-BAO}: Optimized for BAO + CC data, targeting $H_0=73.0$ 
	\item \textbf{F2-SN}: Incorporating Pantheon+ constraints (BAO + CC + SNe), targeting $H_0=69.0$ ($\chi^2$ exceeds $\Lambda$CDM best-fit when targeting $H_0 \gtrsim 69.0$)
\end{itemize} 
The reconstructed $w(z)$ profiles are implemented in \href{http://class-code.net/}{CLASS} with the perturbation parameters ${z_j, c_j, \sigma_j}$ (see Eq. (\ref{eq:delta_w})) fixed to their Fisher-bias derived values.

The results shown in Table \ref{tab:combined_results} and Fig.~\ref{fig:compare_contour} demonstrating robust recovery of target Hubble constants within $1\sigma$ confidence intervals. The F1-BAO model achieved $H_0 = 72.76^{+0.81}_{-0.90}$ from Planck data alone, showing marginal improvement ($\Delta\chi^2 \approx -1.0$) over $\Lambda$CDM for BAO+CC datasets. However, this solution becomes strongly disfavored ($\Delta\chi^2 \approx +40$) when incorporating Pantheon+ constraints, revealing fundamental tensions between supernova distances and the required phantom transition.

\begin{figure*}
	\begin{center}
		\includegraphics[scale = 0.3]{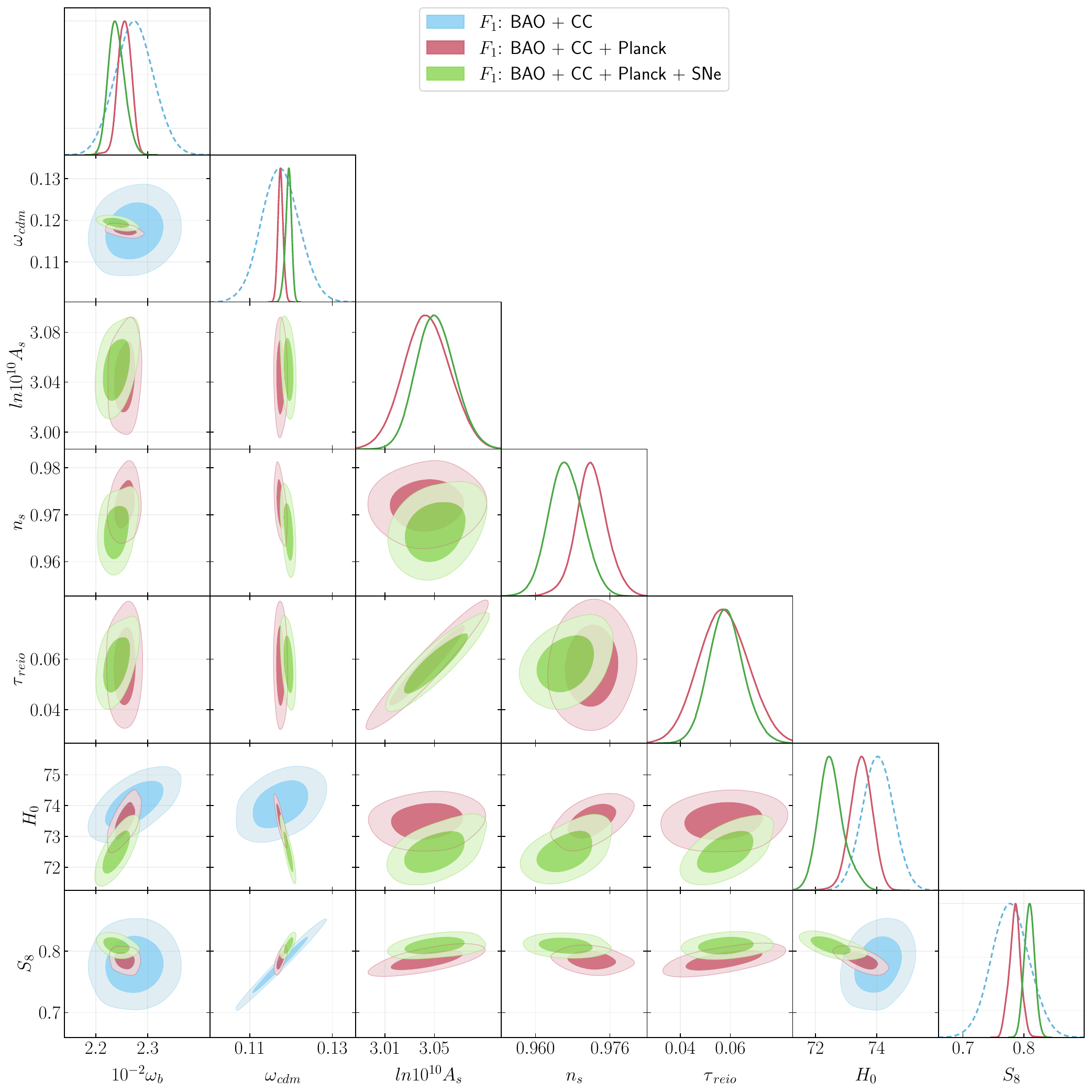}
		\includegraphics[scale = 0.3]{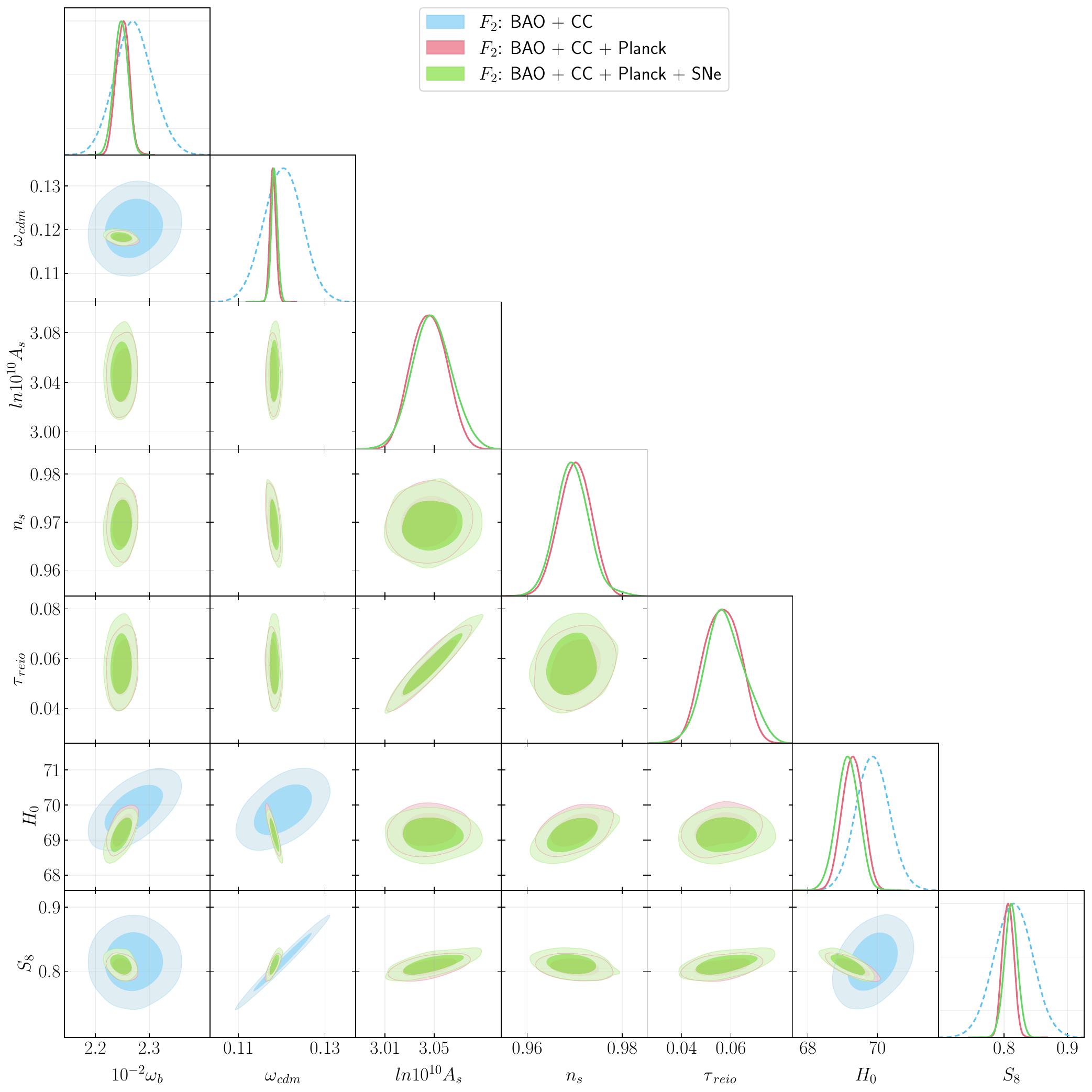}
	\end{center}
	\caption{  Cosmological parameter constraints for the Fisher-bias optimized models from different data combinations. 
		The upper panel shows the F1 model targeting $H_0 = 73.0$, while the lower panel displays the F2 model targeting $H_0 = $. For each model, we show three data combinations: 
		(i) DESI BAO + CC, 
		(ii) BAO + CC + \emph{Planck} , and 
		(iii) BAO + CC + \emph{Planck} + SNe. 
		All contours represent $1\sigma$ and $2\sigma$ confidence regions.
	}\label{fig:compare_contour2}
\end{figure*}

Conversely, the F2-SN model maintains consistent performance across all datasets ($\Delta\chi^2 < 0.1$ as compared to $\Lambda$CDM model) while slightly alleviates the tension with SH0ES measurements without explicit priors. The joint BAO+CC+Planck+SN analysis yields $H_0 = 69.09\pm0.30$  with $\chi^2=1224.73$, consistent with the setting target  $H_0 = 69.0$.
%demonstrating superior compatibility with cosmological probes. This model's success stems from its moderate phantom phase ($w \approx -1.2$ at $z<0.3$) that avoids extreme deviations while preserving agreement with multiple observational constraints. 
The MCMC posterior distributions (see Fig.~\ref{fig:compare_contour2} for the complete parameter constraints.) confirm Gaussian convergence around target parameters, validating the Fisher-bias optimization approach while highlighting Pantheon+'s critical role in constraining viable solutions to cosmological tensions.

\subsection{$S_8$($\Omega_{m,0}$) Tensions}
Both F1-BAO and F2-SN models slightly alleviate the $S_8$ tension between \emph{Planck} and  weak lensing surveys ( DES-Y3 \cite{DES:2021bvc} $S_8 = 0.759^{+0.024}_{-0.021}$) through consistent suppression of late-time structure formation. The mechanism originates from the phantom transition in $w(z)$, which suppresses matter clustering at low redshifts while maintaining compatibility with CMB constraints at earlier epochs.

Notably, the matter density preferences reveal a fundamental tension: whereas Pantheon+ SNe data prefer a higher value $\Omega_{m,0} \approx 0.334$ \cite{Brout:2022vxf}, both Fisher-bias models - particularly F1-BAO with $\Omega_{m,0} = 0.2604\pm 0.0035$ - demonstrate significantly lower matter densities. This discrepancy further corroborates the incompatibility between Pantheon+ constraints and late-time $w(z)$ modifications that successfully reconcile other cosmological tensions, underscoring the unique challenge posed by supernova data in resolving the $H_0$ crisis.

\section{CONCLUSIONS}\label{sec:conclusions}
Our analysis highlights a persistent challenge in addressing cosmological tensions through late-time adjustments to the DE EoS. The Fisher-bias approach shows that while reconstructed $w(z)$ profiles can align BAO+CMB+CC or with local $H_0$ measurements, no unified solution currently satisfies all observational constraints together.

When Pantheon+ supernova data is excluded, the optimized $w(z)$ model can resolve the tension between BAO, CC, \emph{Planck} and SH0ES. A phantom transition around $z\approx0.25$ produces $H_0=73.62\pm0.82$ from Planck data alone, aligning with SH0ES measurements within $0.3\sigma$. This approach also mitigate the $S_8$ tension via suppressed late-time structure formation ($S_8=0.7856\pm0.0096$), while maintaining good agreement ($\Delta\chi^2 < 0$ compared with $\Lambda$CDM) with non-supernova datasets.

The joint analysis of Pantheon+ SNe Ia with neither choice of anchor - the acoustic scale $\theta_*$ at $z\sim1100$ or BAO measurements at $z\sim0.3 -2.0$ - permits solutions with $H_0>69$ while maintaining $\Delta\chi^2\leq0$ relative to $\Lambda$CDM. This reveals that the core tension lies not between SNe, BAO, and CMB measurements collectively, but specifically between the supernova distances and other cosmological probes.
This inconsistency is further corroborated by the matter density parameter $\Omega_m$. The BAO-optimized solution yields $\Omega_m=0.2604\pm0.0035$, while Pantheon+ alone prefers $\Omega_m\approx0.334$ - a significant tension that persists regardless of late-time modifications. The inability to simultaneously reconcile these $\Omega_m$ values while achieving higher $H_0$ demonstrates that supernova distances impose structural limitations on cosmological solutions beyond what can be addressed through expansion history modifications alone.

The above results indicate that resolving current cosmological tensions may demand solutions beyond late-time $w(z)$ adjustments. Potential avenues include combined early- and late-universe physics  or more substantial departures from standard cosmology. 
%Upcoming observations from next-generation surveys will play a key role in evaluating these possibilities and clarifying the nature of these enduring tensions.\\

%between DM and photons on the evolution of primordial
%matter fluctuations and the CMB temperature power spectra.

\begin{acknowledgments}
The Project Supported in part by Natural Science Basic Research Plan in Shaanxi Province of China Program No. 2025JC-YBQN-497 (People's Republic of China).
\end{acknowledgments}

%\acknowledgments

\bibliography{gde}

\end{document}